\documentclass[prb,twocolumn,superscriptaddress]{revtex4-1}

\usepackage{graphicx}
\usepackage{bm}
\usepackage{amsmath}
\usepackage{amsfonts}

\begin{document}


\title{Theory of barrier vs tilt exchange gate operations in spin-based quantum computing}
 \author{Yun-Pil Shim}
 \email{ypshim@lps.umd.edu}
 \affiliation{Laboratory for Physical Sciences, College Park, Maryland 20740, USA}
 \affiliation{Department of Physics, University of Maryland, College Park, Maryland 20742, USA}
 \author{Charles Tahan}
 \email{charlie@tahan.com}
 \affiliation{Laboratory for Physical Sciences, College Park, Maryland 20740, USA}
 \date{\today}

\begin{abstract}
We present a theory for understanding the exchange interaction between electron spins in neighboring quantum dots, either by changing the detuning of the two quantum dots or independently tuning the tunneling barrier between quantum dots. The Hubbard model and a more realistic confining-potential model are used to investigate how the tilting and barrier control affect the effective exchange coupling and thus the gate fidelity in both the detuning and symmetric regimes. We show that the exchange coupling is less sensitive to the charge noise through tunnel barrier control (while allowing for exchange coupling operations on a sweet spot where the exchange interaction has zero derivative with respect to the detuning). Both GaAs and Si quantum dots are considered and we compare our results with experimental data showing qualitative agreements. Our results answer the open question of why barrier gates are preferable to tilt gates for exchange-based gate operations.   
\end{abstract}

\maketitle

\section{Introduction}

Spin qubits in semiconductor quantum dots (QDs) \cite{loss_divincenzo_pra1998,kane_nature1998} are one of the leading candidates for implementation of quantum computing devices \cite{spin_qubit_review,Kloeffel_Loss_annurevcmp2013}. Localized spins such as electron spins in QDs typically have long coherence times since they do not naturally interact with environmental charges. Exchange interactions between spins provide a fast and efficient way of controlling spin-spin interactions and two-qubit gate operations. Engineered local magnetic fields provide a fast electrical control of individual spins by electric dipole spin resonance (EDSR) and individual spin readout by electron spin resonance (ESR) \cite{tokura_tarucha_prl2006,pioro-ladriere_tarucha_nphy2014,kawakami_vandersypen_nnano2014,takeda_tarucha_sciadv2016}. Recent experiments show promising progress towards a scalable architecture for spin qubits \cite{yoneda_takeda_nnano2018,zajac_petta_science2018,vandersypen2017}. 
Another approach for scalable qubit architecture based on localized spins is to encode the qubit states in a two-dimensional subspace of larger systems such as two \cite{Levi_prl2002,Petta2005,Maune_Hunter_nature2012,Shulman_Yacoby_science2012}, three \cite{divincenzo_bacon_nature2000,laird_gossard_prb2010,fong_wandzura_qic2011,Medford_Marcus_nnano2013,Eng_Hunter_sciadv2015}, or four \cite{hsieh_whaley_qip2003} spins, which can expand the types of qubits in QD spin systems. In particular, the three-spin exchange-only qubit allows for full gate operations using only electrical methods controlling the exchange interactions between spins without the need for a g-factor gradient. 

While electrical control of the spin degree of freedom is promising, the exchange interaction reintroduces charge noise to the spin system. In GaAs QDs, nuclear spins of the host material can also be a dominant source of decoherence, but in Si QDs, enriched material can remove most of the spinful nuclear spins leaving spin qubits with very long coherence times \cite{Tyryshkin_Lyon_nmat2012,Saeedi_Thewalt_science2013,Wang_Petta_prl2013,Veldhorst_Dzurak_nnano2014,Muhonen_Morello_nnano2014}. The main decoherence mechanism for Si QD qubit gates is believed to be charge noise \cite{yoneda_takeda_nnano2018}. There is experimental evidence strongly suggesting that the effect of charge noise is local to QDs and charge noise from various sources can be well approximated with fluctuating gate voltages (gate-referred charge noise) \cite{DQD_sweet_spot_Reed_Hunter_prl2016}.
To mitigate the effects of charge noise, there has been recent progress in finding optimal operating points for exchange-based operations\cite{RX_theory_Taylor_prl2013,RX_exp_Medford_prl2013,TQD_sweet_spot_Fei_Friesen_prb2015,TQD_sweet_spot_Burkard_prb2015,DQD_sweet_spot_Reed_Hunter_prl2016,%
DQD_symmetric_operation_Martins_Kuemmeth_prl2016,shim_tahan_prb2016,malinowski_kuemmeth_prb2017}.  
The resonant exchange (RX) qubit \cite{RX_theory_Taylor_prl2013,RX_exp_Medford_prl2013} operates on a partial sweet spot with microwave control for qubit gate operations, and the always-on, exchange-only (AEON) operation regime \cite{shim_tahan_prb2016,malinowski_kuemmeth_prb2017} offers a full sweet spot to idle, one-, and two-qubit operations by tuning the tunnel barrier to control the exchange interaction between spins. 

The main goal of this work is to compare the barrier and tilt exchange gate operations in QD spin devices. There is some lore that tunnel based gates are less sensitive to charge noise than tilting (or "detuning") gates \cite{Hu_DasSarma_Charge_noise_prl2006,Dial_Yacoby_charge_noise_prl2013}, but these beliefs have never been backed up by a rigorous analysis. Recent experiments \cite{DQD_sweet_spot_Reed_Hunter_prl2016,DQD_symmetric_operation_Martins_Kuemmeth_prl2016,zajac_petta_science2018} showed that optimizing the control of exchange interaction can significantly improve the coherence of the operations; they make use of the symmetric operation, which directly controls the tunnel barrier at a sweet spot where the exchange interaction is insensitive to fluctuations in the detuning between QD energy levels up to the first order. This is in contrast to the conventional method of exchange control by tilting the QD energy levels in the detuning regime \cite{Petta2005}.
In this work, we provide a theoretical study on these two different types of exchange control operations between localized spins - barrier or tilt control. We use two different theoretical models: Hubbard model and confining potential model. The Hubbard model is a very simple model with parameters for the QD orbital energies, tunneling, and direct Coulomb interactions, and it provides many fruitful intuitions for QD devices. For example, we can derive analytical expressions for the exchange interaction from the Hubbard model and the sweet spot can be defined as a point where the exchange coupling has a zero derivative with respect to some tuning parameters. But it is not clear a priori how the parameters in the  Hubbard model are affected by charge noise, and it is not straightforward to connect parameters of the Hubbard model with experimental voltage controls on metallic gates used for defining QDs and control electric signals. 
It is usually assumed or implied that the effects of charge noise is mainly on the detuning between QD energy levels, and its effects on the tunneling and Coulomb interactions are often neglected. But there has been no systematic theoretical study on this issue. 
A more general theoretical framework is necessary and for this we use a microscopic confining-potential model where each QD is defined by its confining potential and the barrier potential between QDs. There have been previous work using confining potentials to describe quantum dot devices \cite{li_dassarma_prb2010,yang_dassarma_prb2011,dassarma_prb2011,wang_dassarma_prb2011}, deriving Hubbard model parameters using Heitler-London and Hund-Mulliken approaches. Particularly, recent work by Yang and Wang \cite{yang_wang_pra2017,yang_wang_pra2018} used a similar approach to study the barrier and tilt controls of the exchange coupling in a DQD. Although it provided some qualitative insights on the subject, it used only two s-orbitals and Hubbard model derived by Hund-Mulliken approximation, which significantly underestimates the tunneling between QDs for given confining potentials and neglects some interacting terms that are necessary for quantitative analysis of the system. 
Here we use a microscopic model \cite{Delgado_Shim_prb2007,Shim_Hawrylak_prb2008} that uses the accurate molecular orbitals for the confining potential and full configuration-interaction (CI) calculation. This confining potential model can naturally connect to the voltage controls of the QD devices, making it a useful theoretical tool beyond previous theoretical models based on the Hubbard model.

\begin{figure}
  \includegraphics[width=\linewidth]{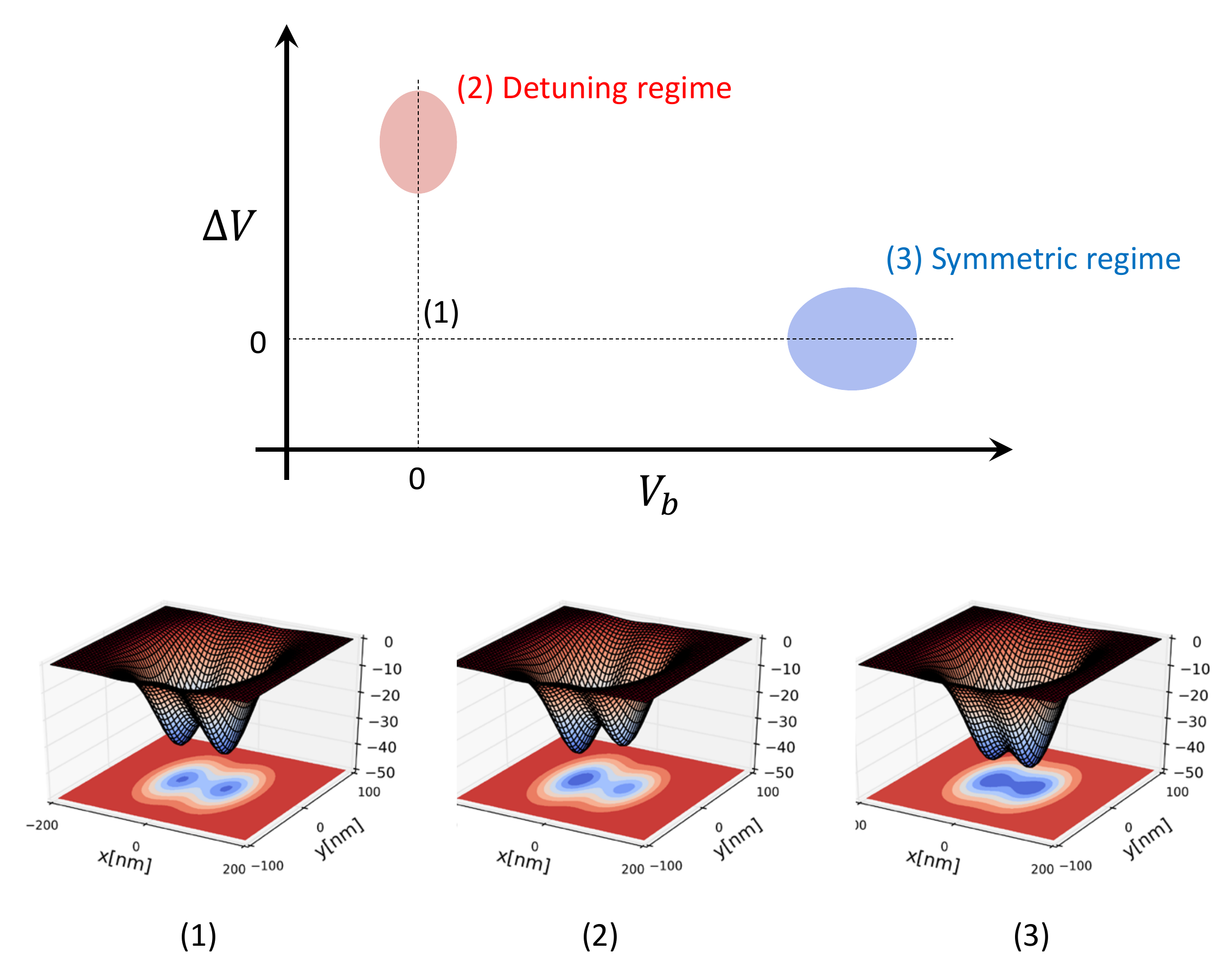}\\
  \caption{Detuning regime and symmetric regime for controlling the effective exchange interaction between electron spins in a DQD. Energy levels in each QD can be shifted by changing the detuning $\Delta V$=$V_1-V_2$ between gate voltages for each QD, and the tunnel barrier is controlled by additional gate voltage $V_b$. Without any detuning or tunnel barrier control, a DQD has well-defined confining potentials for each QD [region (1)]. Control of the effective exchange coupling requires increasing or decreasing the overlap between two QDs. Conventional detuning operation for controlling $J$ is done at large detuning [red shaded region (2)], while symmetric operation uses direct control of the tunnel barrier [blue shaded region (3)]. Lower plots show typical confining potentials of the DQD in each regimes.}
  \label{fig1:Detune_Symm_regimes}
\end{figure}

We consider a double quantum dot (DQD) system and define two regimes of operations in terms of detuning of the DQD. {\it Detuning regime} is where the energy levels of the two QDs are very different (large detuning), and {\it symmetric regime} is where the the energy levels are close (small detuning). A schematic description of these two different regimes are shown in Fig. \ref{fig1:Detune_Symm_regimes}. Conventional detuning operation uses the tilt control in the detuning regime, and the symmetric operation uses the barrier control in the symmetric regime. Detuning operation is easier to implement since only QD orbital energies need to be tuned by gate voltages. Symmetric operation requires an additional control line for the tunnel barrier control, making it more complicated to realize in experiments. Nonetheless, progress in experimental techniques makes this not a critical issue any more, and the symmetric operation was shown to be advantageous over the detuning operation \cite{DQD_sweet_spot_Reed_Hunter_prl2016,DQD_symmetric_operation_Martins_Kuemmeth_prl2016}.     

This paper is organized as follows. In Sec. \ref{sec:hubbard}, we analytically derive the exchange coupling between spins using the Hubbard model and find the sweet spot for the symmetric operation. In Sec. \ref{sec:confining}, we introduce the confining-potential model for multiple QD systems, and presents the results for DQD in Sec. \ref{sec:results}. A summary and further discussion is given in Sec. \ref{sec:conclusion}.

\section{Hubbard model}\label{sec:hubbard}

QD devices are often described by the Hubbard Hamiltonian,
\begin{eqnarray}\label{eq:Hubbard}
\hat{H} 
&=& \sum_{i=1}^{N_Q} \sum_{\sigma} \varepsilon_i c_{i\sigma}^{\dag} c_{i\sigma} 
  + \sum_{i=1}^{N_Q} U_i \hat{n}_{i\uparrow} \hat{n}_{i\downarrow} 
  + \frac{1}{2}\sum_{i\neq j} V_{ij} \hat{n}_i \hat{n}_j \nonumber\\
&& + \sum_{<i,j>} \sum_{\sigma} t_{ij} \left( c_{i\sigma}^{\dag} c_{j\sigma} + h.c. \right)
\end{eqnarray}
where $N_Q$ is the number of QDs, $\varepsilon_i$ is the orbital energy of $i$-th QD, $\hat{n}_i$=$\hat{n}_{i\uparrow} + \hat{n}_{i\downarrow}$, and $\hat{n}_{i\sigma}$=$c_{i\sigma}^{\dag} c_{i\sigma}$. In this section, we use this Hubbard Hamiltonian to derive the standard form of the exchange coupling in terms of the Hubbard model parameters and find conditions for the symmetric operation.


For a system of two electrons in a DQD, we can consider four singly-occupied states $\{ |\uparrow_1,\downarrow_2\rangle, |\downarrow_1,\uparrow_2\rangle, |\uparrow_1,\uparrow_2\rangle, |\downarrow_1,\downarrow_2\rangle \}$ and two doubly-occupied states $\{ |\uparrow_1,\downarrow_1\rangle, |\uparrow_2,\downarrow_2\rangle \}$ as basis states where $\uparrow_i$ ($\downarrow_i$) is the up (down) spin in $i$-th QD. Then the Hamiltonian matrix in this basis is
\begin{widetext}
\begin{equation}
H = \left( 
\begin{array}{cccccc}
\varepsilon_1+\varepsilon_2+V_{12} & 0 & 0 & 0 & t & t \\
0 & \varepsilon_1+\varepsilon_2+V_{12} & 0 & 0 & -t & -t \\
0 & 0 & \varepsilon_1+\varepsilon_2+V_{12} & 0 & 0 & 0 \\
0 & 0 & 0 & \varepsilon_1+\varepsilon_2+V_{12} & 0 & 0 \\
t & -t & 0 & 0 & 2\varepsilon_1+U_1 & 0 \\
t & -t & 0 & 0 & 0& 2\varepsilon_2+U_2 
\end{array}
\right)~,
\end{equation}
\end{widetext}
and after Schrieffer-Wolff transformation\cite{SW_transform_pr1966} we obtain an effective Hamiltonian in the subspace spanned by singly-occupied states
\begin{equation}
H_{\mathrm{eff}} = \left( \varepsilon_1+\varepsilon_2+V_{12} \right) \openone
+ \left( 
\begin{array}{cccc}
-\frac{J}{2} & \frac{J}{2} & 0 & 0 \\
\frac{J}{2} & -\frac{J}{2} & 0 & 0 \\
0 & 0 & 0 & 0 \\
0 & 0 & 0 & 0 
\end{array}
\right)~,
\end{equation}
where
\begin{equation}\label{eq:exchange}
J = 2 t^2 \left( \frac{1}{U_1-V_{12}+\Delta\varepsilon} + \frac{1}{U_2-V_{12}-\Delta\varepsilon} \right) ~.
\end{equation}
The detuning between the two QDs is $\Delta\varepsilon$=$\varepsilon_1-\varepsilon_2$.
This effective Hamiltonian can be written in terms of the spin operators of localized electrons in each QDs,
\begin{equation}
\widehat{H}_{\mathrm{eff}} = E_0 + J \mathbf{s}_1 \cdot \mathbf{s}_2 ~,
\end{equation}
where $E_0$=$ \varepsilon_1+\varepsilon_2+V_{12} -J/4$ and $J$ is the exchange coupling between the two electron spins.
The exchange interaction is a combined effect of the Coulomb interaction and the Pauli exclusion principle, and it allows electrical control of the spin states of the two-spin system. This also introduces the charge noise into the spin states, and it is one of the main decoherence mechanisms in exchange-coupled spin systems. The exchange interaction $J$ can be controlled by tuning the detuning $\Delta\varepsilon$ or the tunneling $t$ or both, by applying gate voltages to the metallic gates. To minimize the effects of charge noise on the exchange operations, it is preferable to work on a sweet spot where the exchange coupling $J$ has vanishing derivative for the tuning parameters.
The derivatives of $J$ with respect to the detuning and the tunneling are 
\begin{eqnarray}
\frac{\partial J}{\partial \Delta\varepsilon} &=& \frac{2t^2 \left(U_1+U_2-2V_{12}\right) \left( 2\Delta\varepsilon + U_1-U_2 \right)}{\left(U_1-V_{12}+\Delta\varepsilon\right)^2 \left(U_2-V_{12}-\Delta\varepsilon\right)^2} ~, \\
\frac{\partial J}{\partial t} &=& 4t \frac{U_1+U_2-2V_{12}}{\left(U_1-V_{12}+\Delta\varepsilon\right)\left(U_2-V_{12}-\Delta\varepsilon\right)} ~.
\end{eqnarray}
For typical QD devices where $U_1, U_2 \gg V_{12}$, we can find a sweet spot w.r.t. the detuning at $\Delta\varepsilon$=$\left( U_2-U_1 \right)/2$, but there is no sweet spot w.r.t. the tunneling $t$ except for the trivial sweet spot $t$=0. Note that, for $U_1$=$U_2$, the sweet spot is at the symmetric point with zero detuning $\Delta\varepsilon$=0. It is therefore preferable to control the tunneling barrier ($t$) while keeping the system on the sweet spot where the exchange coupling $J$ does not change by fluctuations in the detuning (due to charge noises) up to first order. But the charge noise also affects the tunneling $t$ and the Coulomb interaction $U_1$, $V_{12}$, and we need a method beyond Hubbard model to take this into account. Hubbard model provides a simple and intuitive model for the QD devices, but when comparing with experiments, Hubbard parameters ($\varepsilon_i$, $t$, $U_i$, $V_{ij}$) are typically fitted to the experimental data using some phenomenological models, and it is not straightforward to connect with experimental controls of various gate voltages.

\section{Confining-potential model}\label{sec:confining}

In this section, we remedy the shortcomings of the Hubbard model by starting with a model that can simulate the experiments reasonably, while keeping it simple enough not to require full device simulation such as tight-binding calculations. In this confining-potential model, each QD is defined by its confining potential. We chose Gaussian potentials to describe both the quantum dot confining potentials and the barrier potential, which is more realistic than truncated harmonic potentials.  
First, we calculate the molecular orbitals of a single electron in a DQD. The Hamiltonian of an electron in the DQD is given by
\begin{equation}
\widehat{H}_{\mathrm{DQD}} =  \frac{\mathbf{p}^2}{2 m} + V_{\mathrm{DQD}} \left( \mathbf{r} \right) + V_{\mathrm{barrier}} \left( \mathbf{r} \right) ~,
\end{equation}
where  $m$ is the effective mass of the electron and the DQD confining potential is 
\begin{equation}
V_{\mathrm{DQD}} \left( \mathbf{r} \right) = \sum_{j=1}^2 - V_{j} \exp\left[-\frac{\left(\mathbf{r}-\mathbf{r}_{0,j}\right)^2}{d_{j}^2} \right] ~,
\end{equation} 
and the barrier potential is
\begin{equation}
V_{\mathrm{barrier}} \left( \mathbf{r} \right) = - V_{b} \exp\left[-\frac{\left(\mathbf{r}-\mathbf{r}_{b}\right)^2}{d_{b}^2} \right]~.
\end{equation}
$\mathbf{r}_{0,j}$ is the position of $j$-th QD and $\mathbf{r}_{b}$ is the position of the barrier between the two QDs.
The Gaussian confining potential is approximately a harmonic potential near the potential minimum with harmonic-oscillator (HO) frequency $\omega_j$=$\sqrt{2 V_{0,j}/m d_j^2 }$.
We use the Fock-Darwin (FD) orbitals \cite{Kouwenhoven_review} of HO potentials as the basis states, and find the molecular orbitals as linear combinations of the HO orbitals (LCHO) by solving the eigenvalues and eigenstates of $\widehat{H}_{\mathrm{DQD}}$.
The shape of the confining potential is determined by $V_{0,j}$ and $d_{j}$, which can be independently controlled in principle. 
In this work, for the sake of simplicity, we change $V_{0,j}$ while keeping $\Omega_j$=$\hbar\omega_j/Ry$ fixed where $Ry$=$me^4/2\epsilon^2\hbar^4$ is the effective Rydberg constant that we choose as the energy scale. We will also use the effective Bohr radius $a_0$=$\epsilon\hbar^2/me^2$ as the length scale. Here $\epsilon$ is the dielectric constant, and $-e$ is the electronic charge.   
We consider both GaAs and Si QDs. For GaAs $Ry$=5.93meV, $a_0$=9.79nm, and for Si $Ry$=25meV, $a_0$=2.5nm.

Using the LCHO molecular orbitals, the Hamiltonian for multiple electrons is 
\begin{eqnarray}\label{eq:many-body-H}
\widehat{H} &=& \sum_{k,\sigma} \varepsilon_k c^\dag_{k\sigma} c_{k\sigma} \nonumber \\
            &+& \frac{1}{2}\sum_{\substack{k_1,k_2,k_3,k_4,\\ \sigma,\sigma'}} \langle k_1 k_2 | \widehat{V} | k_3 k_4 \rangle 
                 c^\dag_{k_1\sigma} c^\dag_{k_2\sigma'} c_{k_4\sigma'} c_{k_3 \sigma}  ~,
\end{eqnarray} 
where $k$'s are indices for LCHO orbitals and $\varepsilon_k$ is the energy of the $k$-th molecular orbital. The Coulomb matrix element is
\begin{equation}
\langle k_1 k_2 | \widehat{V} | k_3 k_4 \rangle = \int d\mathbf{r}d\mathbf{r}' 
    \frac{\phi^*_{k_1} (\mathbf{r}) \phi^*_{k_2} (\mathbf{r}') \phi_{k_3} (\mathbf{r}) \phi_{k_4} (\mathbf{r}')}{4\pi\epsilon_0 \epsilon | \mathbf{r}-\mathbf{r}' |}  ~,
\end{equation}
where $\phi_{k} (\mathbf{r})$ is the wavefunction of the $k$-th molecular orbital.
We perform the full configuration-interaction (CI) calculation to find the many-electron states \cite{Delgado_Shim_prb2007,Shim_Hawrylak_prb2008}.
The effective exchange interaction $J$ between electron spin in each QDs is defined as the energy difference between the lowest triplet and singlet states, $J$=$E_T-E_S$.
In this model, the changes in $V_j$'s and $V_b$ naturally simulate the voltage changes of metallic gates defining each QDs and the tunnel barrier between QDs, respectively. To obtain reasonable convergence, we used 45 FD orbitals from each QDs (90 FD orbitals in total) to calculate the molecular orbitals of the DQD, and then used 30 lowest-energy molecular orbitals to perform the CI calculation with two electrons in the DQD.

\section{Numerical results}\label{sec:results}

Now we present results from numerical simulations of DQD devices using confining-potential model discussed above.  
Following the experiments \cite{DQD_sweet_spot_Reed_Hunter_prl2016,DQD_symmetric_operation_Martins_Kuemmeth_prl2016}, we set the distance between the two QDs to be about 100nm. So we set $\mathbf{r}_{0,1}$=$(-5, 0)$, $\mathbf{r}_{0,2}$=$(5, 0)$ in unit of $a_0$ for GaAs, and $\mathbf{r}_{0,1}$=$(-20, 0)$, $\mathbf{r}_{0,2}$=$(20, 0)$ for Si. The barrier potential is in the middle of the two QDs, $r_b$=$(0,0)$ First we tried to find reasonable values for the parameters $V_j$'s, $d_j$'s, $V_b$, and $d_b$ to get the effective exchange coupling $J$ in the range of $\mu$eV, and then fine-tuned the parameters to simulate the barrier and tilt operations for controlling the exchange coupling $J$.

\subsection{Detuning operation vs Symmetric operation}

\begin{figure}
  \includegraphics[width=\linewidth]{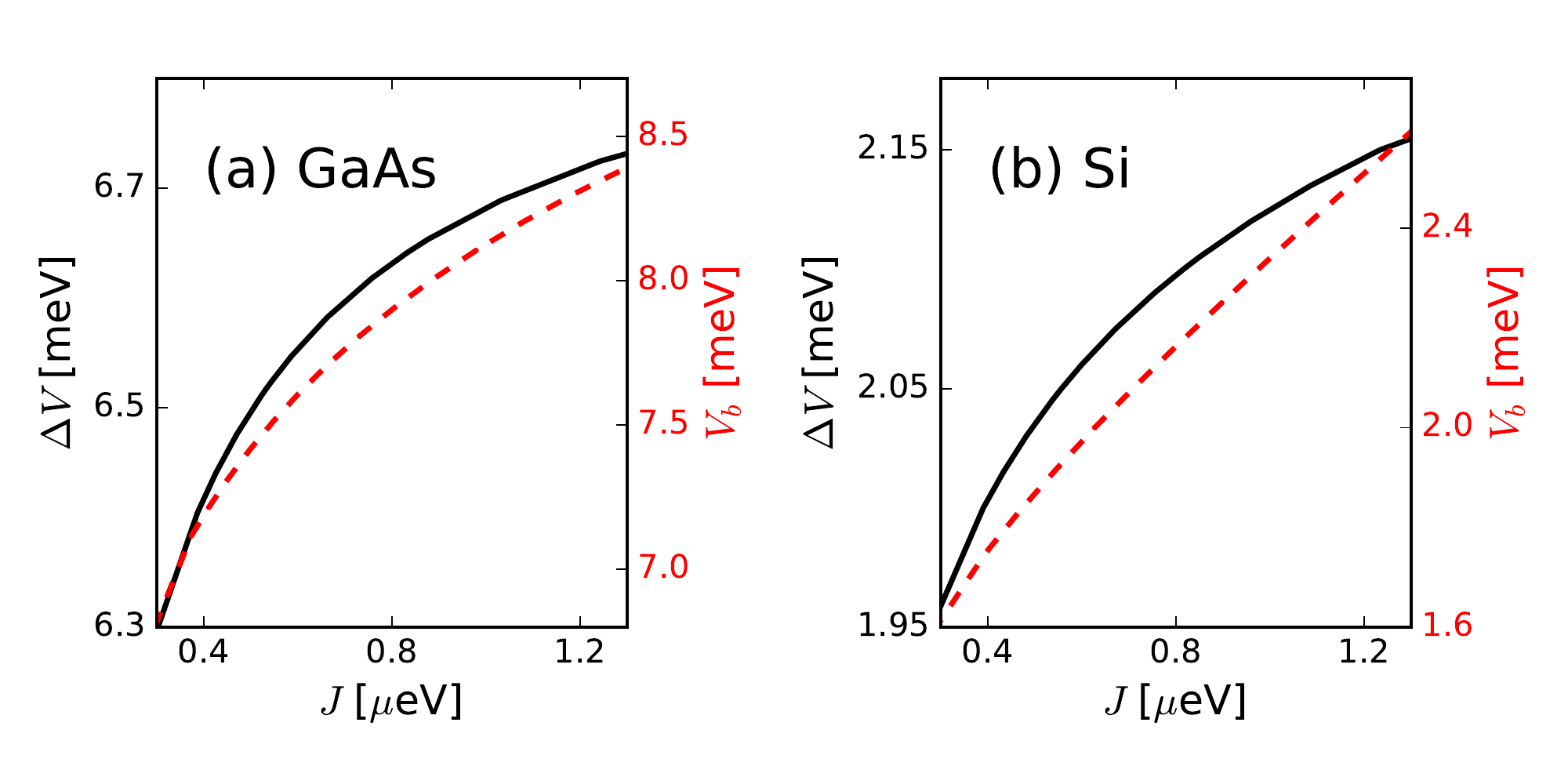}\\
  \caption{Detuning operation vs Symmetric operation. Solid black curves are for detuning operations and dashed red curves are for symmetric operations. For a similar range of $J$ controls (0.3 to 1.3 $\mu$eV), symmetric operation needed larger change in $V_b$ compared with the change in $\Delta V$ in detuning operations. This is the case for both GaAs and Si devices, and agrees with experiment \cite{DQD_symmetric_operation_Martins_Kuemmeth_prl2016} in GaAs DQD. These results indicate that the tunneling is less sensitive to charge noise than the detuning.}
  \label{fig2:detune_vs_symm}
\end{figure}

Tilting the potential energies of the DQD allows for easy control over the effective exchange interaction, by tuning the effects of the virtual double occupation of one of the dots. On the other hand, one can directly control the tunnel barrier to increase/decrease the overlap between QDs.
We consider these two types of control of $J$: {\it Detuning operation} uses tilt control in the detuning regime (regime (2) in Fig. \ref{fig1:Detune_Symm_regimes}) and {\it symmetric operation} uses barrier control in the symmetric regime (regime (3) in Fig. \ref{fig1:Detune_Symm_regimes}). In Fig. \ref{fig2:detune_vs_symm}, we numerically calculated the effective exchange coupling $J$ while changing $\Delta V$=$V_1-V_2$ (energy difference between confining potential minima of QDs) for detuning operation or changing $V_b$ (barrier potential minima) for symmetric operation.

For the detuning operation in GaAs (Fig. \ref{fig2:detune_vs_symm}(a), solid black curve), we fix $\Omega_1$=$\Omega_2$=$2\sqrt{6}/5$. 
The barrier potential was set to be zero ($V_b$=0). Then we change the detuning $\Delta V$ while keeping the average $(V_1+V_2)/2$=6$Ry$ fixed. To tune $J$ in the range between 0.3 $\mu$eV and 1.3 $\mu$eV, $\Delta V$ was needed to be changed by 0.4meV (from 6.3meV to 6.7meV). 

For the symmetric operation in GaAs (Fig. \ref{fig2:detune_vs_symm}(a), dashed red curve), we fix $V_1$=$V_2$=6$Ry$ and $d_1$=$d_2$=5$a_0$. Then we changed $V_b$ while keeping $\Omega_b$=$2\sqrt{6}/5$. For the same range of tunability for $J$ as in the detuning operation, we need to change $V_b$ by about 1.5meV. 
Since this operation is being done at the symmetric point ($\Delta V$=0), $\partial J/\partial \Delta V$=0 and the charge noise on $\Delta V$ does not affect $J$ to the lowest order. In addition, since the barrier control requires larger range of change in $V_b$ compared to the change in $\Delta V$ in detuning operation (by about a factor of 4 in this case), the effect of charge noise on $V_b$ is expected to be less severe than its effects on $\Delta V$ in detuning operation. This result agrees with the experimental data for GaAs DQD in Ref. \onlinecite{DQD_symmetric_operation_Martins_Kuemmeth_prl2016} and confirms that the effects of charge noise on the tunnel barrier in the symmetric regime would be much less significant than its effects on the detuning in the detuning regime.   

We did similar numerical simulation of the DQD in silicon. For the detuning operation in Si (Fig. \ref{fig2:detune_vs_symm}(b), solid black curve), $\Omega_1$=$\Omega_2$=$\sqrt{0.1}/10$ which corresponds to $V_1$=$V_2$=0.1$Ry$ and $d_1$=$d_2$=20$a_0$. 
We could get the exchange coupling $J$ in the similar range as in GaAs by changing $\Delta V$ from 1.95 to 2.15meV. For the symmetric operation in Si (Fig. \ref{fig2:detune_vs_symm}(b), dashed red curve), keeping $V_1$=$V_2$=0.1$Ry$ and $d_1$=$d_2$=20$a_0$, $V_b$ was changed by 0.9meV (from 1.6 to 2.5meV) to achieve the same range of $J$. Similar to the GaAs case, we need to change $V_b$ in symmetric operation more than $\Delta V$ in detuning operation by more than 4 times. Therefore we expect less sensitivity for charge noise in symmetric operation than in detuning operation.     
We could also see that this is true for smaller J values in general although the sensitivity difference ratio decreases somewhat.

Note that the Hubbard model cannot predict this type of different behavior of the tilt and barrier controls, while this confining-potential model correctly predicts that the barrier control is less sensitive to the charge noise.
 
\subsection{Tilt control in symmetric regime}

\begin{figure}
  \includegraphics[width=\linewidth]{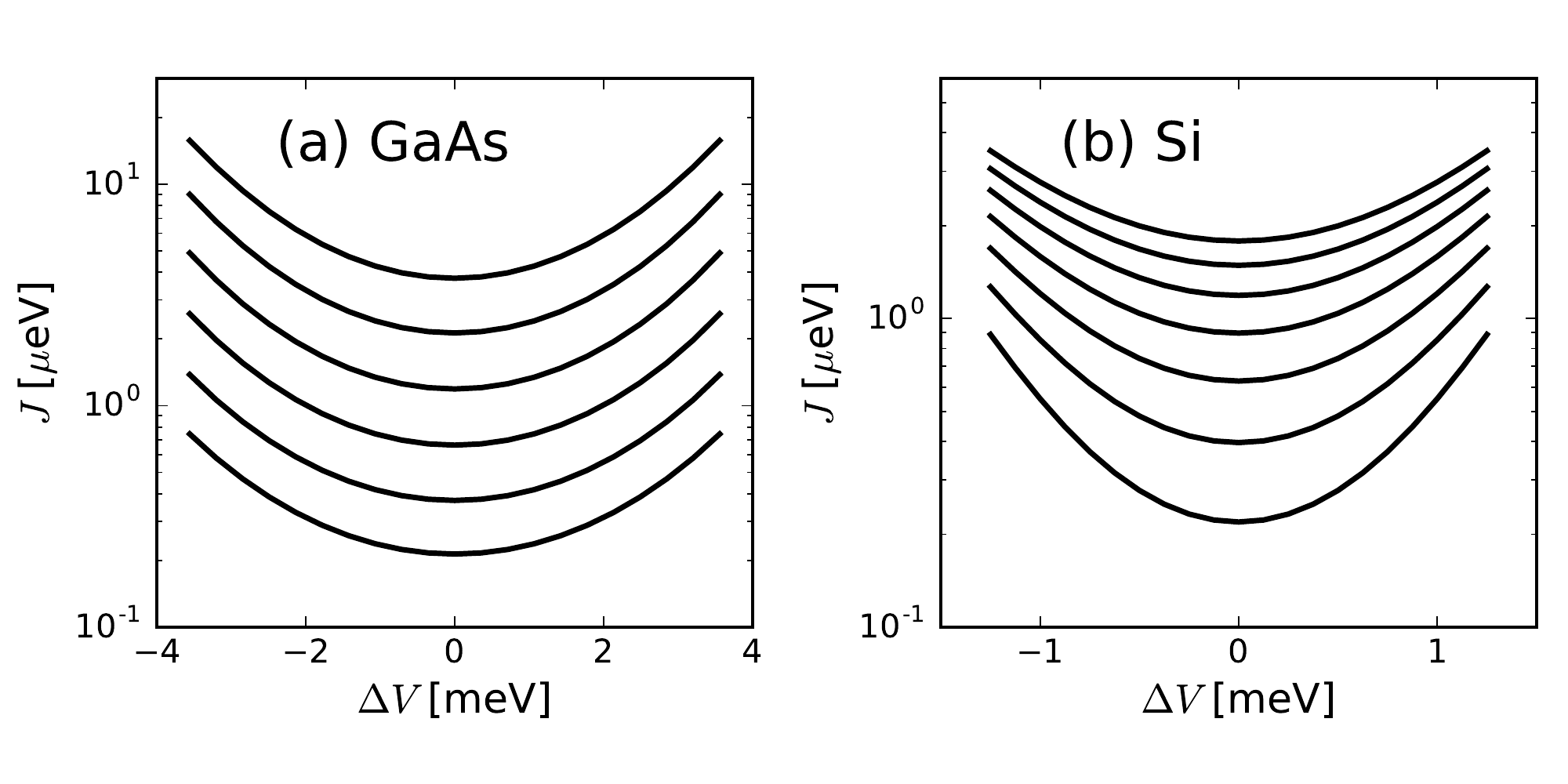}\\
  \caption{Tilt control in symmetric regime. Each curve has fixed $V_b$ and we changed $\Delta V$. For GaAs in (a), from top to bottom, $V_b$=1.6, 1.5, 1.4, 1.3, 1.2, 1.1 $Ry$. For Si in (b), from top to bottom, $V_b$=0.12, 0.11, 0.10, 0.09, 0.08, 0.07, 0.06 $Ry$. We can see the sweet spot at zero detuning, and we need relatively large change in $\Delta V$ to achieve similar control over $J$ compared with in detuning regime.}
  \label{fig3:tilt_in_symm}
\end{figure}

We also considered tilt control of the exchange coupling in symmetric regime. In this case, the barrier-control voltage is kept at a value to tune $J$ in a desired range (e.g. $\mu$eV range), then tilting control is used to fine-tune $J$. Figure \ref{fig3:tilt_in_symm} shows how $J$ changes as a function of $\Delta V$. Each curves correspond to different values of $V_b$. First, we can see that the exchange interaction is minimum at exact symmetric point ($\Delta V$=0) as it should. This is the sweet spot for symmetric operation. Second, to achieve similar tunability in $J$ (e.g. from 0.3 $\mu$eV to 1.3 $\mu$eV) as in Fig. \ref{fig2:detune_vs_symm}, the change in $\Delta V$ we need is much larger than in detuning operation and is comparable to the change in $V_b$ in the symmetric operation. This means that the charge noise on $\Delta V$ in the symmetric regime is not as damaging as in the detuning regime. But this tilt control necessarily moves the device out of the sweet spot making it more vulnerable to charge noise.

\subsection{Barrier control in detuning regime}

\begin{figure}
  \includegraphics[width=\linewidth]{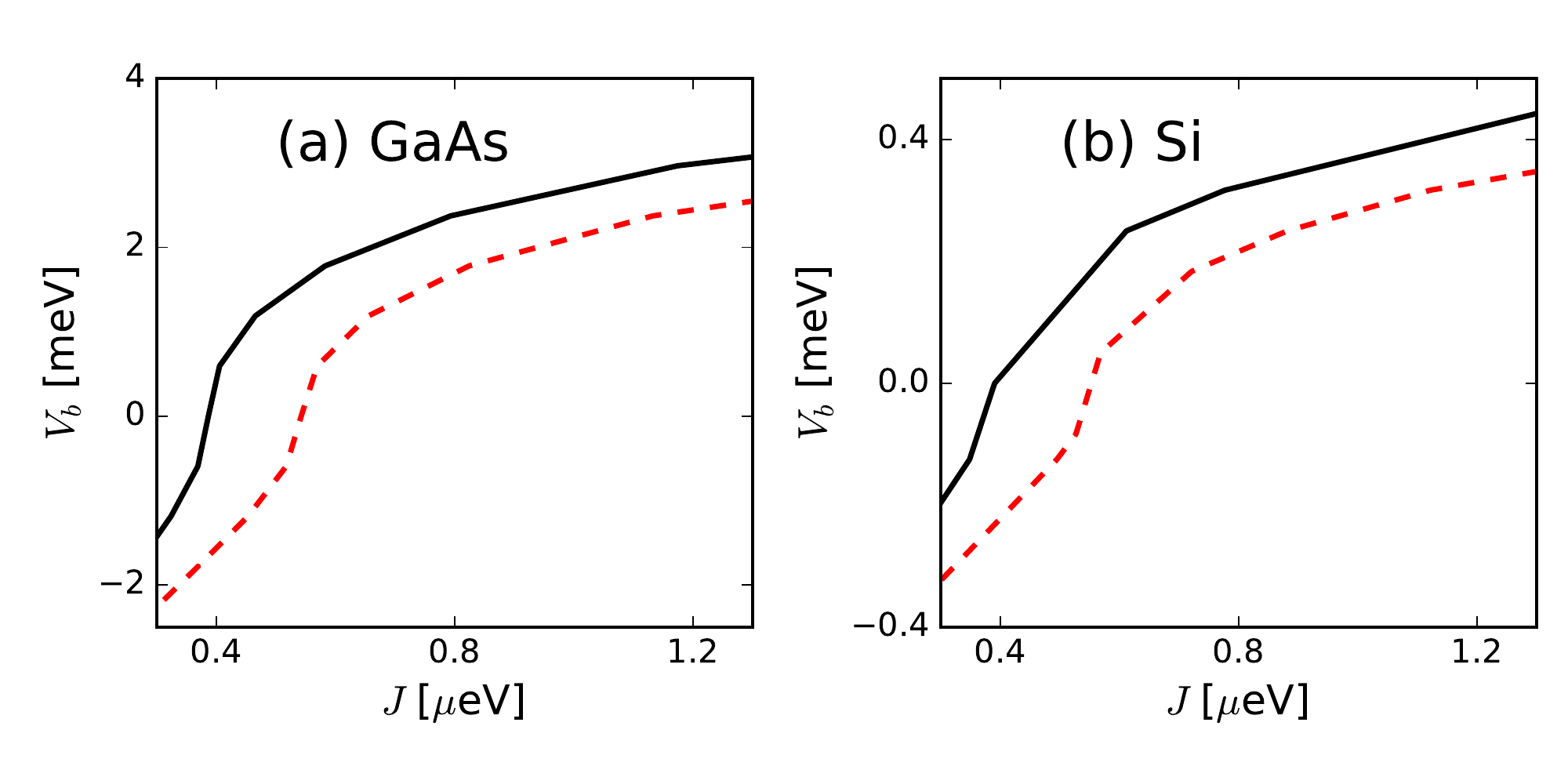}\\
  \caption{Barrier control in detuning regime. For GaAs in (a), solid black curve is for $\Delta V$=1.08$Ry$ and dashed red curve is for $\Delta V$=1.10$Ry$. For Si in (b), solid black curve is for $\Delta V$=0.08$Ry$ and dashed red curve is for $\Delta V$=0.082$Ry$. In this detuning regime, a slight change in $\Delta V$ results in rather large change in $J$, making it more vulnerable to charge noise in $\Delta V$. For the barrier controls in these detuning regimes, we needed to change $V_b$ significantly more to achieve similar range of $J$ compared to previous cases.}
  \label{fig4:barr_in_detune}
\end{figure}

Lastly, we consider tuning the barrier potential in detuning regime where the detuning $\Delta V$ is set so that $J$ is in $\mu$eV range, then we control the barrier voltage $V_b$ to fine-tune the exchange interaction $J$. In Fig. \ref{fig4:barr_in_detune}, we show the change in $J$ as we change $V_b$ in detuning regime. Slightly different values of $\Delta V$ can cause quite large change in $J$, as can be seen by comparing the solid black and dashed red curves. For both GaAs and Si devices, large changes in $V_b$ are needed to achieve $J$-control in the same range as in previous cases. Therefore, in detuning regimes, the charge noise in $\Delta V$ causes large fluctuations in $J$ while $J$ is relatively much more robust against the charge noise in the tunneling barrier, implying that barrier control could be better in the detuning regime too.

\subsection{Global energy shift}

\begin{figure}
  \includegraphics[width=\linewidth]{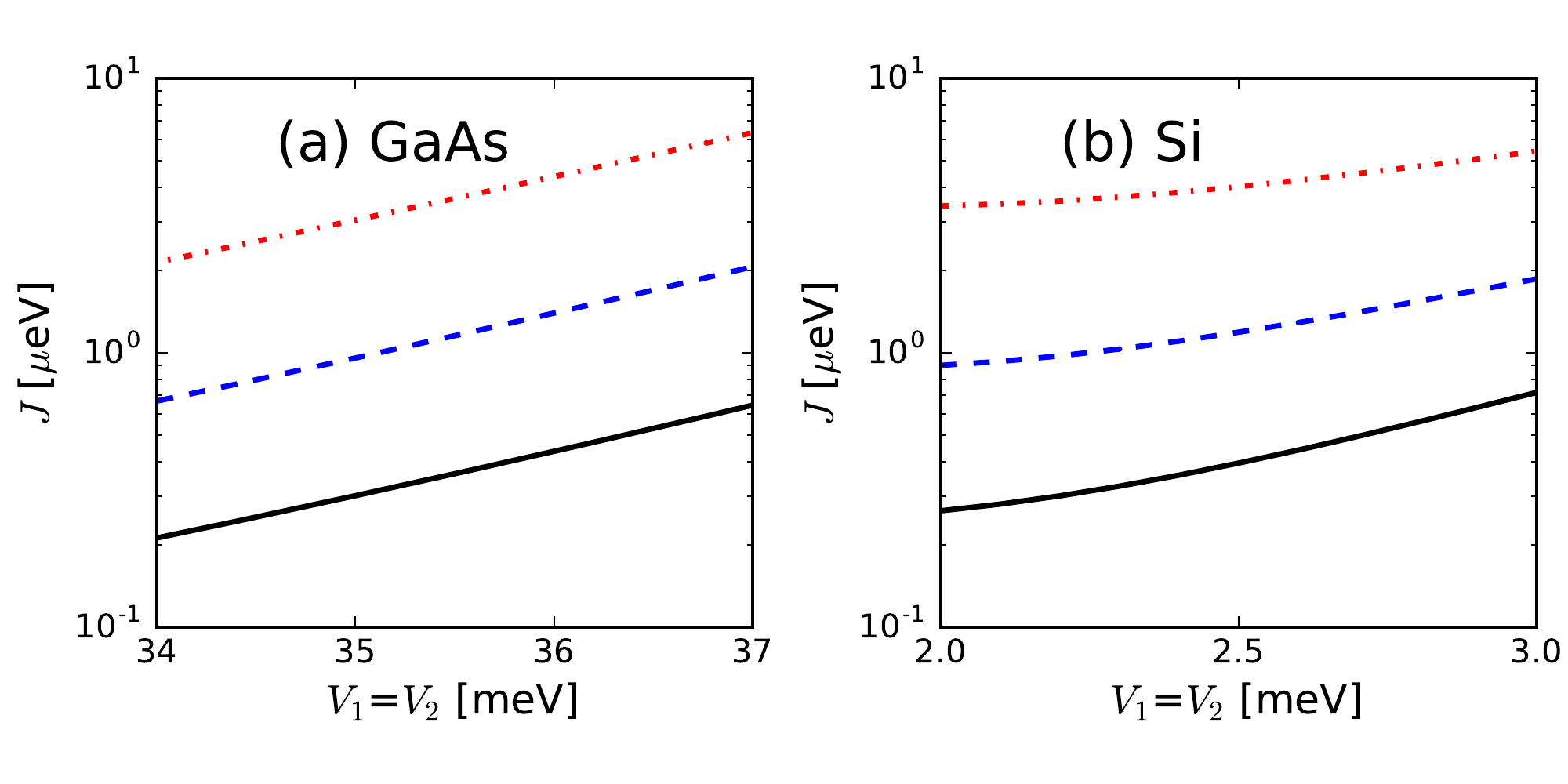}\\
  \caption{Change of the exchange interaction $J$ when both $V_1$ and $V_2$ are shifted in the same way. Different curves correspond to different values of $V_b$. For GaAs in (a), $V_b$=1.2$Ry$ for solid black curve, 1.4$Ry$ for dashed blue curve, and 1.6$Ry$ for dot-dashed red curve. For Si in (b), $V_b$=0.07$Ry$ for solid black curve, 0.1$Ry$ for dashed blue curve, and 0.22$Ry$ for dot-dahsed red curve. In all cases, we can see that $J$ does not change much with this global shift.}
  \label{fig5:global_shift}
\end{figure}
In the Hubbard model, the global energy shift of QD energy levels does not affect the exchange interaction since the Hubbard parameters are not affected. But in realistic models such as the confining-potential model, the global shift of the gate voltages should also change the exchange interaction since it changes the shape of the confining potential ( although its effect should be much smaller than changes in the detuning). In Fig. \ref{fig5:global_shift}, we show how the exchange coupling $J$ changes as we change both $V_1$ and $V_2$ in the same way (i.e. $V_1$=$V_2$). Even for changes of a few meV in $V_1$, the exchange coupling $J$ does not change significantly as was expected. This justifies that the effects of changes in the global shift are negligible.

\section{Conclusion}\label{sec:conclusion}

We provide a relatively simple theoretical framework that can describe the exchange gate operations in various regimes for QD devices.
Particularly, barrier vs tilt control of the exchange interaction are systematically studied showing that the insensitivity on the gate charge noise for barrier control in symmetric regime can potentially lead to efficient implementation of quantum gates with higher fidelities than the conventional tilting gates in the detuning regime. A further study including explicit noise model would be needed to quantify the effectiveness of both types of gate controls. 
We also show that barrier gates are better for mitigating charge noise even in the detuning regime. In GaAs, nuclear spins provide another noise mechanism that can be dominant over the charge noise for some ranges of the exchange interaction \cite{DQD_symmetric_operation_Martins_Kuemmeth_prl2016}, and it is more complicated since the advantage of the symmetric operation in the gate fidelities could depend on the magnetic noise in the system \cite{zhang_dassarma_prl2017}. For Si, charge noise is the dominant decoherence source after isotopically enriching the host material for $^{28}$Si\cite{yoneda_takeda_nnano2018}. Therefore, symmetric operation via barrier control will be an essential ingredient in scalable spin qubit architecture as was demonstrated for high fidelity single- and two-qubit gates \cite{zajac_petta_science2018}. 

Symmetric operation will also be critical in encoded qubit schemes since they are mostly based on exchange gate operations. In fact, the triple QD exchange-only qubit \cite{shim_tahan_prb2016} where all qubit logical gate operations can be done by only using exchange interactions while keeping the qubits on a full sweet spot can be considered as a generalization of the symmetric operation to triple QD system.

\begin{figure}
  \includegraphics[width=\linewidth]{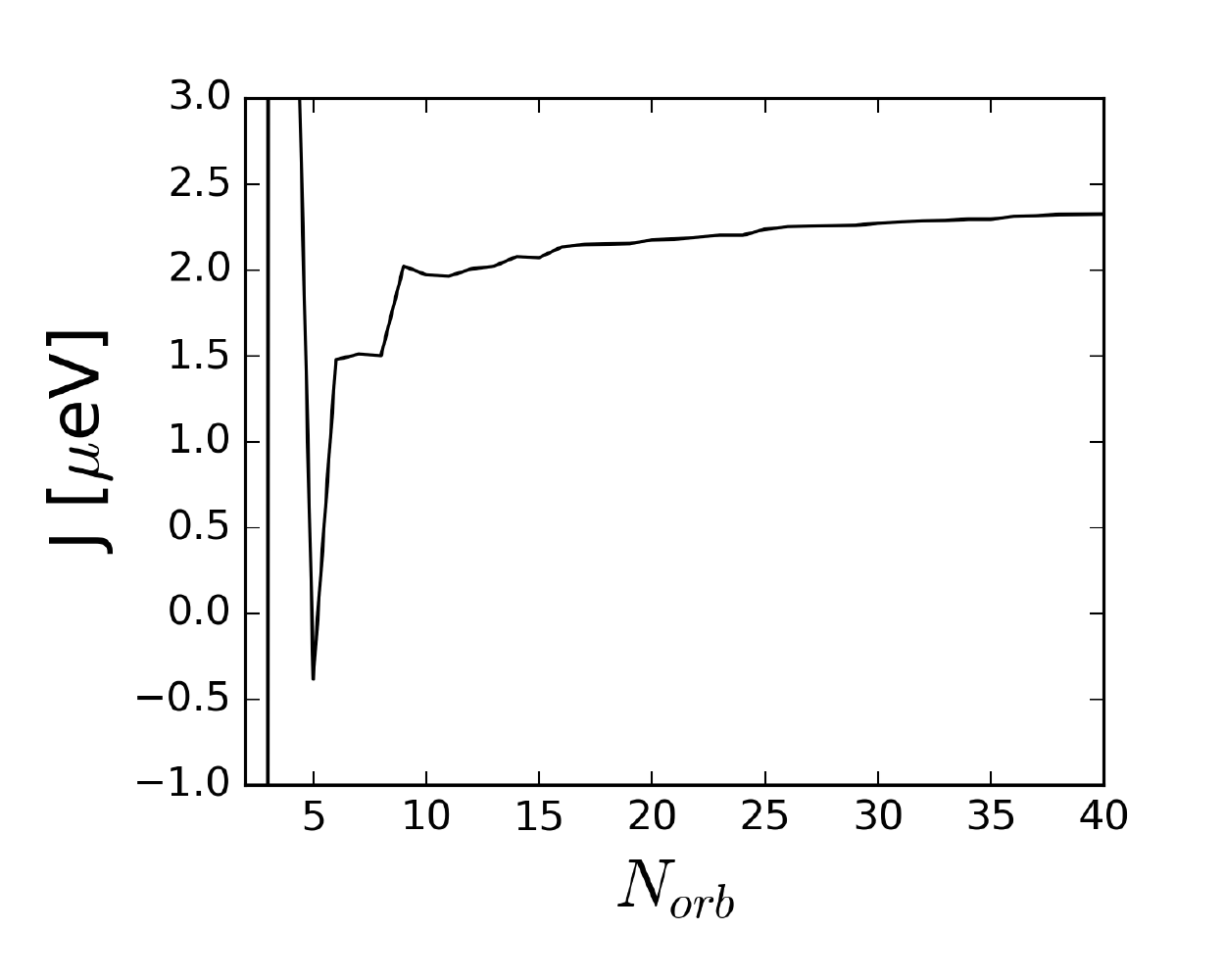}\\
  \caption{Exchange coupling $J$ calculated using varying number of molecular orbitals $N_{orb}$ in CI. It requires tens of molecular orbitals to achieve convergence.}
  \label{fig6:convergence}
\end{figure}

The confining-potential model presented here is relatively simple. Note that here we did not try to fit the confining potential shapes to the experimental gate structures. Simple Gaussian forms for the confining potentials and the barrier potential were used, and we tried to find appropriate values for $V_i$'s and $d_i$'s to get similar values of $J$ as in the experiments. So we cannot directly compare the numerical results with experimental data. But even without such full device simulation, we successfully reproduce most qualitative experimental results for GaAs DQD (Ref. \onlinecite{DQD_symmetric_operation_Martins_Kuemmeth_prl2016} and its Supplementary Material) and predicts similar behavior for Si DQD. 
Simpler models used in previous works are quite limited in quantitative analysis since it typically involves very limited number of orbitals.
To achieve reasonably good convergence, it is necessary to include many molecular orbitals in the CI calculation. Figure \ref{fig6:convergence} shows an example of the calculated effective exchange coupling $J$ as a function of the number of molecular orbitals used in CI. 
We used 30 molecular orbitals in this work to assure good convergence while keeping compuational demand manageable.
 
This tool should be useful for theoretically studying multi-QD devices more realistically. These analyses show that it is preferable to put fast lines on the barrier gates and use them for operations while filtering the dot gates as much as possible to maximize gate fidelity.


\bibliographystyle{apsrev4-1}

%

\end{document}